# WHITE-BOX MEMBERSHIP ATTACK AGAINST MACHINE LEARNING BASED RETINOPATHY CLASSFICATION

**Mounia Hamidouche (1, 2), Reda Bellafqira (2), Gwenolé Quellec (1), Gouenou Coatrieux (2)**

*1. Université de Bretagne Occidentale - UBO, Inserm, UMR 1101, 29238 Brest, France; 2. Institut Mines-Telecom, IMT Atlantique, Inserm, UMR 1101, 29238 Brest, France*

## Introduction

The advances in machine learning (ML) have greatly improved AI-based diagnosis aid systems in medical imaging. However, being based on collecting medical data specific to individuals induces several security issues, especially in terms of privacy. Even though the owner of the images like a hospital put in place strict privacy protection provisions at the level of its information system, the model trained over his images still holds disclosure potential. The trained model may be accessible to an attacker as: **1) White-box:** accessing to the model architecture and parameters; **2) Black-box:** where he can only query the model with his own inputs through an appropriate interface. Existing attack methods include: feature estimation attacks (FEA), membership inference attack (MIA), model memorization attack (MMA) and identification attacks (IA) [1]. In this work we focus on MIA against a model that has been trained to detect diabetic retinopathy from retinal images. Diabetic retinopathy is a condition that can cause vision loss and blindness in the people who have diabetes. MIA is the process of determining whether a data sample comes from the training data set of a trained ML model or not. From a privacy perspective in our use case where a diabetic retinopathy classification model is given to partners that have at their disposal images along with patients' identifiers, inferring the membership status of a data sample can help to state if a patient has contributed or not to the training of the model.

## Methodology

MIAs have been increasingly studied since the work in [3] who performed the first membership attack in the context of genome-wide association studies. MIAs have been further performed in many different settings using various types of models and data sets [4]. In this work, we follow the methodology proposed in [5] and evaluate it in the white-box setting to retinal data set. Working under the white-box hypothesis and in the context of retinopathy are the two main contributions of our work. The driven solution for MIA exploits the fact that ML models behave differently with respect to their training data compared to data they have not seen, i.e., over-fitting. As depicted in Fig. 1, the membership inference attack we are focusing on in its original version relies on: a Shadow model, the purpose of which is to make a copy of the victim model (a priori unknown from the attacker) taking inputs different from the training set; a Binary classifier, that is trained on the posteriors obtained from the shadow models and their corresponding membership labels, in order state if an input has been used to train the shadow models. The basic idea of this attack is to say that if the binary discriminates well training inputs of the shadow model, it will have the same behavior for the training inputs of the victim model.

The MIA approach proposed here has been implemented considering the victim model as a diabetic retinopathy classification model. Its purpose is to predict which class does the given fundus image belongs to; classes herein defined as: normal, mild, moderate, severe, proliferative. Having a direct access to the victim model parameters we apply this attack as follows:

**Step 1: Shadow models:** as we consider white-box setting, our shadow model architecture is similar to the victim model architecture. More clearly, the shadow model is a fine-tuned version of the victim model with a different training set since the attacker does not have access to the original training set, by definition. Notice however that these data are taken from the same population as the training data.

**Step 2: Binary classifier:** At the end of the shadow models training, we get all the posterior probability outputs with the train images labeled "in" and the test images labeled "out". Then, based on these outputs, we perform a binary classification to predict whether an image comes from the training data set.

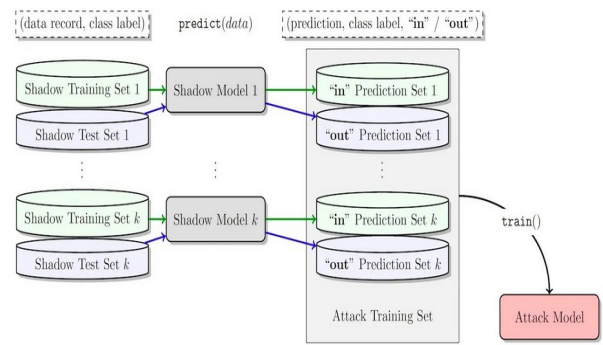

Fig. 1 Membership inference attacks observe the behavior of a target machine learning model and predict examples that were used to train it [5].

## Experiments

In this section, we evaluate the attack described above on retinal images taken from kaggle[1]. It consists of 3262 images from 5 classes we divide into a training set of 1500 images for the target model, a set of 462 images for testing and a last set of 1300 samples to train the shadow models. The presented experiment has been deployed as follows:

**1) Training of a diabetic retinopathy classifier** - We have used two pre-trained models: ResNet18 and EfficientNet5. We set the learning rate to 0.001, and the maximum epochs of training to 80. We obtained an accuracy of 73,62% on the training set and 72,36% on the test set with ResNet18 architecture; whereas with EfficientNet5, we obtained an accuracy of 96,64% on the training set and 76,81 on the test set (over-fitting).

**2) Shadow models construction** - Under the white-box hypothesis scenario, we use one shadow model with the same architecture as the victim model. Then, we train the shadow model with a data set of retinal images different from those used to train the victim model.

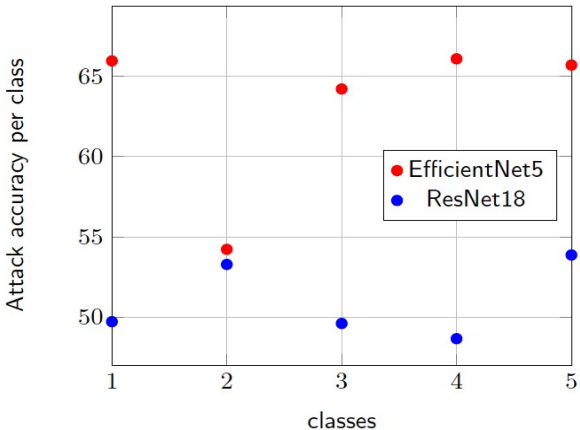

Fig. 2 Precision of the membership inference attack on each class of the victim model.

**3) Binary classifier training:** Given as input the posterior probability and the membership label of each sample obtained from 2) we train a support vector machine (SVM) to learn whether a sample is member of the training data set or not.

The attack performance is quantified using the standard precision, i.e., the fraction of the records inferred as members of the training data set that are indeed members. The accuracy of the attack is 51,42% when the shadow model is trained with ResNet18, and 60,03% with EfficientNet5. Considering the fact that an attack precision around 50% is equivalent to random guessing, our results show that the attack is effective when the victim model is over-fitting. The measurements are also reported per class because the accuracy of the attack can vary considerably from a class to another. In Fig. 2 we show the precision of the MIA on each class when the shadow model is trained with ResNet18 and EfficientNet5, respectively. Indeed, our experiments reveal that the attack accuracy attains 66% for the classes 1 and 4 when the shadow model is trained with EfficientNet5. This is mainly due to the difference in size and composition of the training data belonging to each class. However, when the model is trainned using ResNet18, the attack accuracy does not change compared with the over all attack accuracy.

---

[1] https://www.kaggle.com/c/diabetic-retinopathy-detection/overview

## Conclusion

We evaluated the effectiveness of MIA in the context of machine learning based retinopathy detection. Even though we worked on simple ML models, the obtained results show that over-fitting is a necessary condition for the attack to be efficient as demonstrated in previous works. A possible future work would be to improve the membership attack technique for models that generalize well for our use case.